\documentclass[conference]{IEEEtran}

\IEEEoverridecommandlockouts

\ifCLASSINFOpdf

\else

\fi

\usepackage{cite}
\usepackage{amsmath,amssymb,amsfonts}
\usepackage[compact]{titlesec}

\titlespacing{\section}{12pt}{*0}{*0}

\usepackage[brazil]{babel}
\usepackage{algorithmic}
\usepackage{graphicx}
\usepackage{textcomp}
\usepackage{xcolor}
\usepackage{subfigure}
\usepackage{hyperref}

\begin{document}

\title{Comparando Estratégias de Roteamento em Redes Quânticas Oportunísticas}

\author{\IEEEauthorblockN{Diego Abreu, Alan Veloso, Antônio Abelém}

\thanks{
    Diego Abreu, Universidade Federal do Pará (e-mail: \url{diego.abreu@itec.ufpa.br})
}

\thanks{
    Alan Veloso , Universidade Federal do Pará (e-mail: \url{aveloso@ufpa.br})
}

\thanks{
    Antônio Abelém, Universidade Federal do Pará (e-mail: \url{abelem@ufpa.br})
}
  
}


\maketitle

\begin{abstract}
Este artigo apresenta uma análise comparativa de três estratégias de roteamento em redes quânticas oportunísticas. As redes de comunicação quânticas enfrentam desafios únicos, como a fragilidade dos qubits e a necessidade de criar e manter pares de estados emaranhados para transmissão confiável. Nesse contexto, o roteamento eficiente e confiável é crucial para maximizar a fidelidade das rotas estabelecidas, minimizar a criação de novos pares de emaranhados e reduzir a necessidade de recálculo de rotas. As estratégias de roteamento são comparadas com base na fidelidade das rotas escolhidas, na quantidade de pares de emaranhados criados e no número de recálculos das rotas. Os resultados obtidos fornecem informações valiosas para o projeto e otimização de redes quânticas oportunísticas, contribuindo para avanços na eficiência e confiabilidade das comunicações quânticas.
\end{abstract}


\IEEEpeerreviewmaketitle

\section{Introdução}
A comunicação quântica vem avançando nos últimos anos em com novas tecnologias de computadores quânticos e  diversas aplicações sendo propostas \cite{abelem2023quantum}.  No entanto, diferentemente das redes de computadores atuais, as redes quânticas enfrentam desafios únicos devido à natureza dos qubits (bits quânticos), que são os blocos de construção dos sistemas quânticos. Assim, para garantir a transmissão confiável de informações quânticas, é necessário estabelecer rotas eficientes e confiáveis. Nesse contexto, o roteamento desempenha um papel crucial, permitindo a seleção das melhores rotas entre os nós de uma rede quântica, considerando as limitações físicas e as restrições impostas pelo meio ambiente quântico \cite{wquantum}.

Em redes quânticas oportunísticas \cite{farahbakhsh2022opportunistic}, onde os qubits são enviados assim que pares de estados emaranhados (EPRs) se tornam disponíveis ponto a ponto, o roteamento assume uma importância ainda maior. A eficiência na escolha das rotas afeta diretamente a fidelidade dos estados quânticos transmitidos, impactando a qualidade e a confiabilidade da comunicação quântica. Além disso, uma rota mal selecionada pode resultar na criação excessiva de novos pares de EPRs, que consomem recursos preciosos e podem comprometer o desempenho geral da rede. Portanto, é fundamental comparar e avaliar estratégias de roteamento para identificar aquelas que melhor atendam às necessidades das redes quânticas oportunísticas.

Neste contexto, este artigo apresenta uma análise comparativa de três estratégias de roteamento em redes quânticas oportunísticas. O objetivo é investigar e avaliar o desempenho das diferentes estratégias em termos de fidelidade das rotas estabelecidas, quantidade de pares de EPRs criados e necessidade de recálculo das rotas. Os resultados obtidos fornecerão \textit{insights}  para o projeto e otimização de redes quânticas oportunísticas, contribuindo para compreender e aprimorar o roteamento em redes quânticas, o que é essencial para impulsionar a adoção da computação quântica e comunicação segura.

\section{Modelagem de  Redes Quânticas Oportunísticas}
A rede quântica pode ser modelada como um  grafo $G = (V, E) $, onde $V$ é o conjunto de nós quânticos e $E$ é o conjunto de arestas que conectam os nós. Cada nó quântico $i$ $\epsilon$ $V$ possui propriedades associadas, representadas por um conjunto de parâmetros. Essa modelagem é feita de forma que cada nó no grafo representa um nó quântico, ou seja, um local onde os qubits são armazenados e processados. Por sua vez, as arestas do grafo representam os canais de comunicação entre os nós quânticos. Cada aresta é responsável por transmitir qubits e tem propriedades associadas, como a disponibilidade de pares de EPRs, fidelidade do canal e dinâmicas de redução e reposição da fidelidade. Essas propriedades refletem as características do canal de comunicação, como a capacidade de transmitir informações quânticas com alta fidelidade e a necessidade de criar e manter EPRs para garantir a confiabilidade da comunicação.


\vspace{-.5cm}
\begin{figure}[!htb]
  \centering
    \includegraphics[width=.2\textwidth]{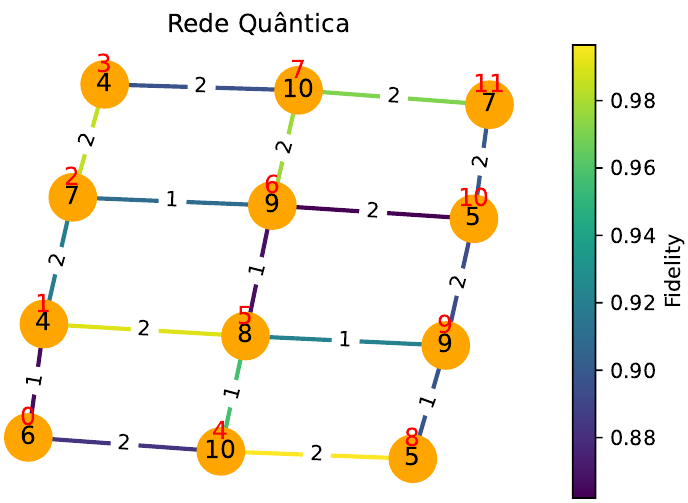}
    \vspace{-.2cm}
  \caption{Grafo representando uma rede quântica de 12 nós, com topologia em treliça (\textit{lattice)}.}
  \label{propostasbrc22}
\end{figure}

\vspace{-.4cm}

\begin{figure*}[h]  
  \begin{minipage}[h]{0.3\linewidth}
    \centering
    \includegraphics[scale=0.32]{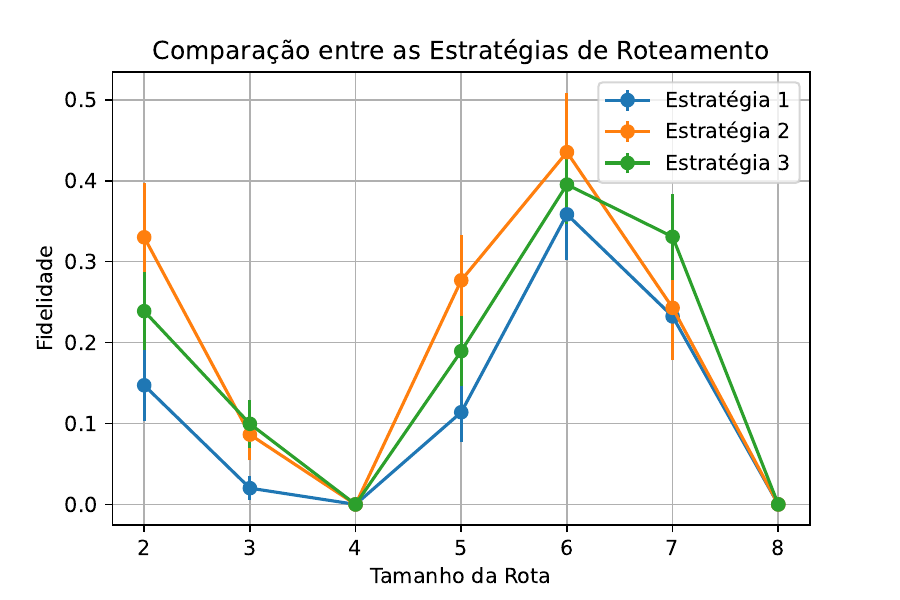}
    \vspace{-.2cm}
    \caption{Fidelidade fim-a-fim.}
    \label{fig:resultado_fidelidade_g1}
  \end{minipage}%
  \hspace{0.1cm}
  \begin{minipage}[h]{0.3\linewidth}
    \centering
    \includegraphics[scale=0.32]{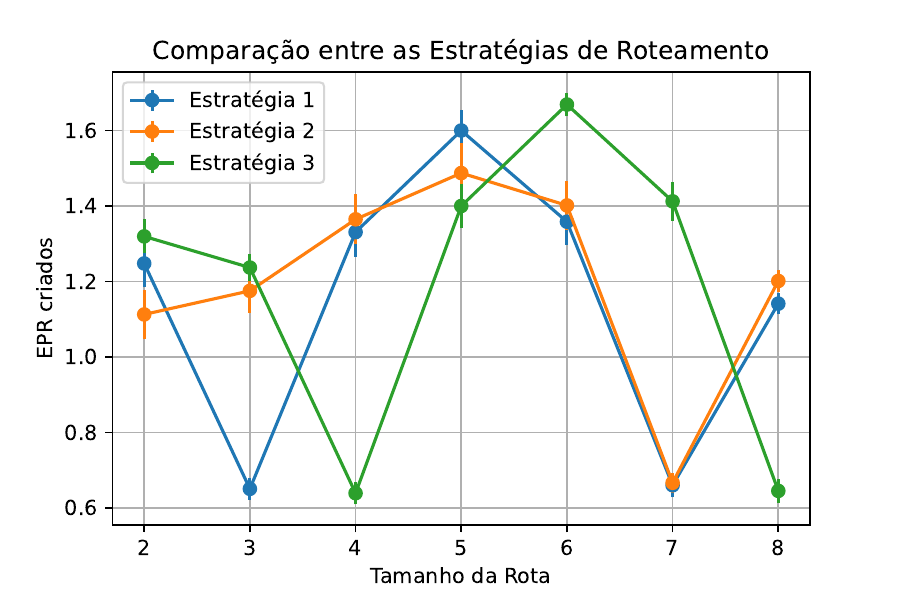}
    \vspace{-.2cm}
    \caption{EPR criados por qubit enviado.}
    \label{fig:resultado_EPR_g1}
  \end{minipage}
    \hspace{0.1cm}
   \begin{minipage}[h]{0.3\linewidth}
    \centering
     \includegraphics[scale=0.32]{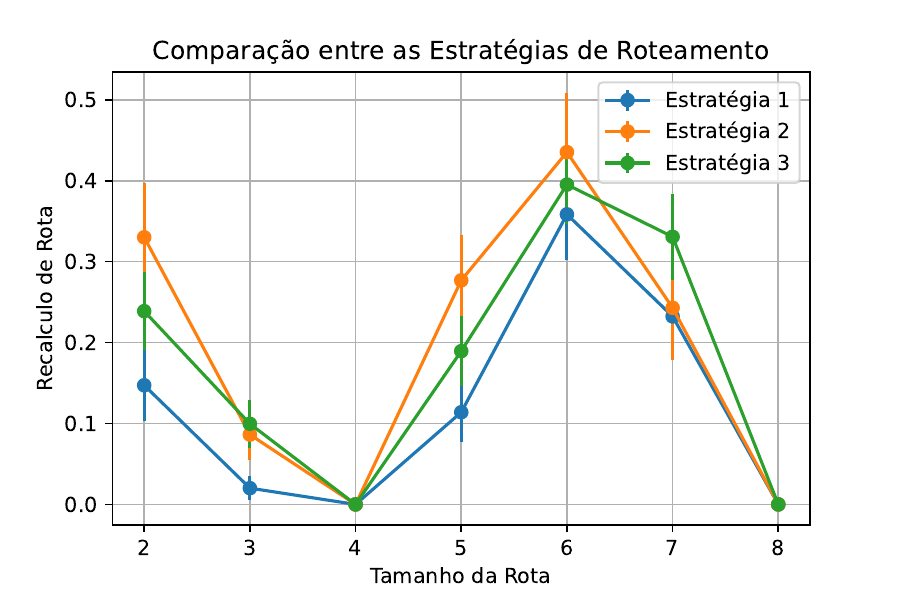}
     \vspace{-.2cm}
     \caption{Recalculo de rota por qubit enviado.}
     \label{fig:resultado_reca_g1}
  \end{minipage}
  \vspace{-.5cm}
\end{figure*}  

\begin{figure*}[h]  
  \begin{minipage}[h]{0.3\linewidth}
    \centering
    \includegraphics[scale=0.32]{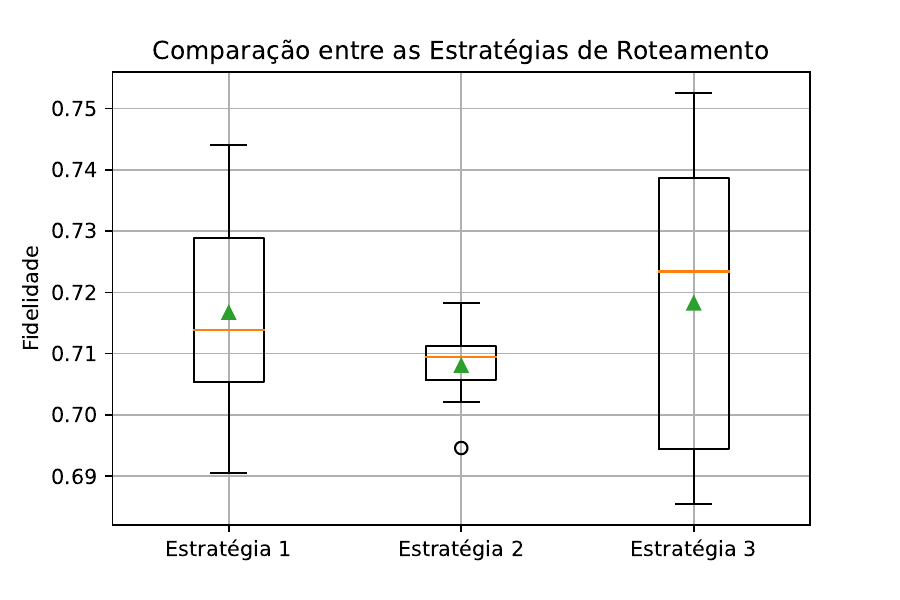}
    \vspace{-.2cm}
    \caption{Fidelidade fim-a-fim.}
    \label{fig:resultado_fidelidade_g2}
  \end{minipage}%
  \hspace{0.1cm}
  \begin{minipage}[h]{0.3\linewidth}
    \centering
    \includegraphics[scale=0.32]{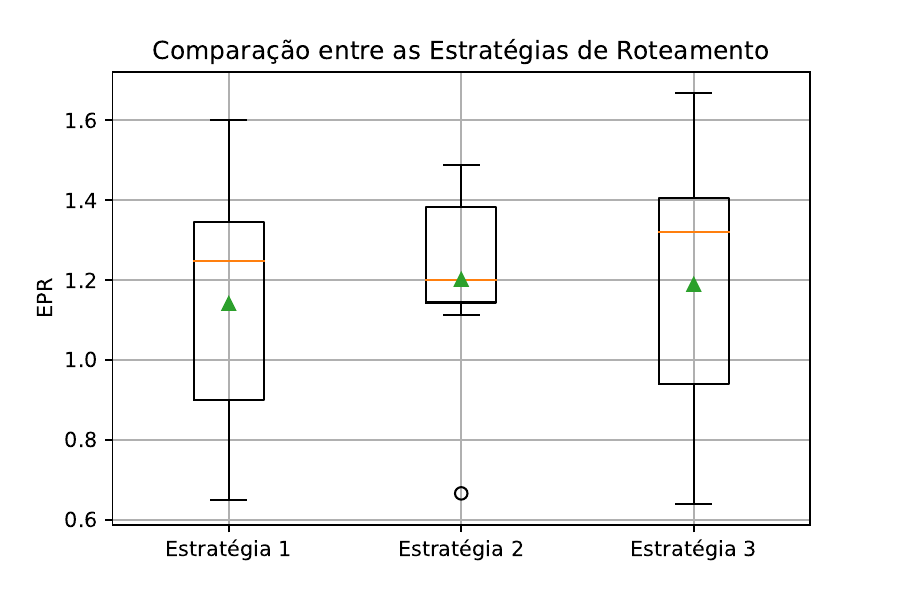}
    \vspace{-.2cm}
    \caption{EPR criados por qubit enviado.}
    \label{fig:resultado_epr_g2}
  \end{minipage}
    \hspace{0.1cm}
   \begin{minipage}[h]{0.3\linewidth}
    \centering
     \includegraphics[scale=0.32]{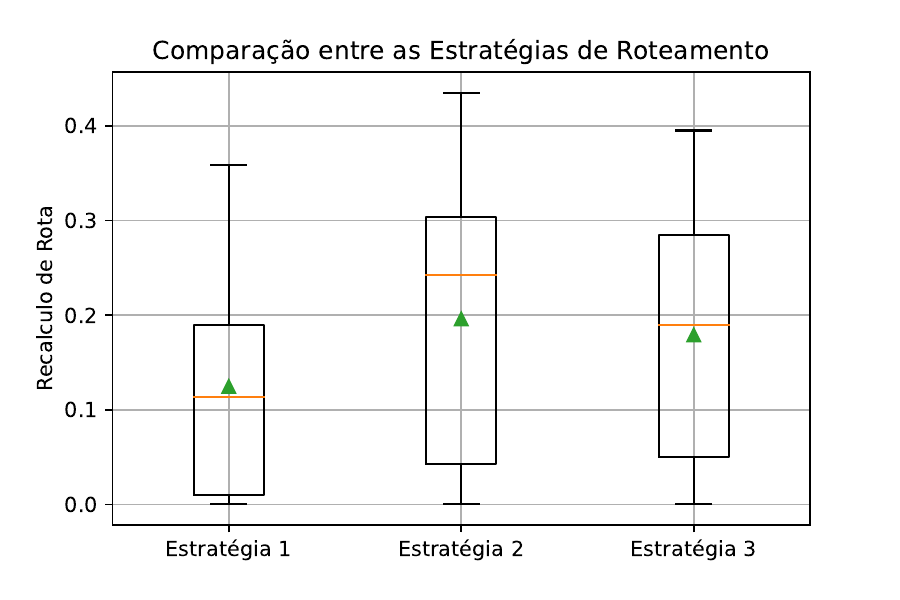}
     \vspace{-.2cm}
     \caption{Recalculo de rota por qubit enviado.}
     \label{fig:resultado_recaL_g2}
  \end{minipage}
  \vspace{-.5cm}
\end{figure*} 

A Fig. \ref{propostasbrc22} apresenta um exemplo de grafo que modela uma rede quântica de 12 nós, com topologia em treliça, em um dado momento da rede. Cada nó tem um índice (em vermelho) e possui uma determinada quantidade de qubits (escrito nos círculos). Cada canal tem uma quantidade de EPR (escrito nas arestas) e uma fidelidade (indicada pela cor das arestas seguindo a legenda de gráfico de cores).

Para se enviar um qubit de um nó para outro na rede é necessário utilizar o protocolo de teletransporte quântico. Utilizando 1 par EPR a cada salto da rota, o qubit é enviado e sua qualidade depende da fidelidade fim-a-fim da rota escolhida. Um par EPR só pode ser utilizado uma vez, após isso ele é consumido e precisa ser novamente criado. Para se criar um par EPR é necessário utilizar 2 qubits, um de cada nó do canal. Assim, além da fidelidade,  é necessário sempre monitorar a existência de par EPR no canal para se escolher a rota. Outro fator a se considerar é a decorrência dos qubits e variação da fidelidade do canal. A cada interação da rede, como envio de qubit ou criação de par EPR, o valor do qubits disponível e a fidelidade do canal podem ter sido reduzidas. Isso impacta na qualidade da rota escolhida inicialmente, podendo acarretar em recálculos de rota \cite{wquantum}.

Em redes EPR tradicionais para se enviar um qubit pelo protocolo de teletransporte quântico, é necessário que cada canal na rota tenha um par EPR estabelecido dedicado para essa operação. Se algum canal não tiver um par EPR, é necessário esperar até que esse par seja criado. Isso se torna um grande desafio para a escala de redes quânticas. Assim, a estratégia de redes quânticas oportunísticas, que envia o qubit ponto-a-ponto, sem que toda rota fim-a-fim precise estar estabelecida se torna uma opção mais viável para o roteamento quântico \cite{wquantum}.

Nesse trabalho são consideradas três estratégias de roteamento quântico em redes oportunísticas. Todas elas buscam encontrar a melhor rota fim-a-fim que melhor satisfaça a métrica escolhida. Estratégia 1: Encontra a rota de maior fidelidade fim-a-fim. Estratégia 2: Encontra a rota com maior quantidade de par EPR disponível. Estratégia 3: Encontra a rota com mais Qubits disponível.

\section{Resultados}

Para se demostrar o funcionamento da rede e realizar a comparação entre as estratégias de roteamento, foi modelado e simulado o envio de 100 qubits na rede, e coletando métricas como a fidelidade fim-a-fim média da rota final, a quantidade média de EPR que tiveram que ser criados a cada qubit enviado e a quantidade de vezes que a rota teve que ser recalculada por qubit enviado. Essa operação foi simulada 100 vezes e os valores médios foram coletados junto com o desvio padrão. Além disso, esse procedimento foi repetido realizado variando-se o tamanho da rota escolhida de 2 até 8.
As Fig.\ref{fig:resultado_fidelidade_g1}, \ref{fig:resultado_EPR_g1} e \ref{fig:resultado_reca_g1} apresentam resultados das 3 estratégias de roteamento para as métricas indicadas, variando-se o tamanho da rota. As Fig. \ref{fig:resultado_fidelidade_g2}, \ref{fig:resultado_epr_g2} e \ref{fig:resultado_recaL_g2} apresentam o \textit{boxplot} para os resultados obtidos. 
Os resultados indicam que a Estratégia 2 possui uma fidelidade fim-a-fim média maior do que as outras estratégias, ao variar-se o tamanho da rota. No entanto, essa estratégia resulta em um maior número de recálculos de rota. Quanto à criação de EPR, as Estratégias 1 e 3 apresentam os melhores resultados, variando de acordo com o tamanho da rota. Em resumo, os resultados ressaltam a importância de avaliar os \textit{trade-offs} entre fidelidade fim-a-fim e número de recalculos de rota, bem como escolher a estratégia mais adequada para a criação de EPR, considerando o tamanho da rota específica em cada contexto.

\section*{Agradecimentos}
Agradecemos ao apoio da Fundação de Amparo a Pesquisa do Estado de São Paulo (FAPESP), por meio do processo no 2020/04031-1.

\section{Conclusão e Trabalhos Futuros}
Nesse trabalho foi realizado a comparação entre 3 estratégias de roteamento em redes quânticas oportunísticas. Os resultados indicam que dependendo da configuração da rede uma estratégia ou outro pode ser mais adequada. 
Como trabalhos futuros, pretende-se testar estratégias mais robustas, que consideram o custo de cada operação, além de outras topologias de rede mais complexas e com maior numero de nós.


\bibliographystyle{IEEEtran}
\bibliography{references}
\end{document}